\documentstyle[preprint,eqsecnum,aps,floats]{revtex}
\begin{document}

%
\begin{titlepage}
\title{Born--Infeld theory of electroweak and gravitational fields: Possible
correction to Newton and Coulomb laws}
\author{Dmitriy Palatnik\thanks{SEL, Northwestern University Library,
Evanston, IL 60208-3530 US; palatnik@ripco.com}
}

\maketitle

\tighten
\begin{abstract}

In this note one suggests a possibility of direct observation of the
$\theta$-parameter, introduced in the Born--Infeld theory of electroweak
and gravitational fields, developed in quant-ph/0202024. Namely,
one may treat $\theta$ as a universal constant, responsible for correction to
the Coulomb and Newton laws, allowing direct interaction between electrical
charges and masses.

\end{abstract}

\end{titlepage}

\tighten

1. In the note \cite{0202024} a Born--Infeld (BI) \cite{BI} theory of
electroweak
and gravitational fields with the action invariant under $\theta$-rotations
of the Lorentz group, imposed on pairs ($\gamma_a,\, \pi_a$), was developed.
The following equations (slightly modified\footnote{{ }%
Here definition for the energy is $E = c \int \overline\psi p^0 \psi
d\Sigma_0$,
and for the electrical charge is $Q = q \int \overline\psi \gamma^0
\psi d\Sigma_0$,
where $\psi$ is a spinorial wave function of a particle, and $q$ is
electrical charge of electron
(absolute value).}
from (6.5) and (6.6) in \cite{0202024}) were obtained:
\begin{eqnarray}\label{E}
E' &=& \cosh\theta\, E + {{\hbar c}\over{\ell q}}\,\sinh\theta\, Q\,;\\
\label{Q}
Q' &=& \cosh\theta\, Q + {{\ell q}\over{\hbar c}}\, \sinh\theta\,E\,.
\end{eqnarray}
Here
\begin{eqnarray}
\label{ell}
\ell^2 & = & {{10}\over{3\alpha}}(1 + 2\sin^2\theta_W)\,L_p^2\,,
\end{eqnarray}
where $L_p$ is the Planck's length, $\alpha$ is the fine structure constant
for electron, and $\theta_W$ is the Weinberg mixing angle \cite{Halzen}.
Equations (\ref{E}) and (\ref{Q}) mean that if ($E$, $Q$) are possible
solutions of field equations following from the theory for the energy
and the electrical charge of, say, an electron, then ($E'$, $Q'$) are
also possible
solutions for the energy and the electrical charge of the same electron.

2. Consider, now, nonrelativistic case, so that $E \approx mc^2$, and
$E' \approx m'c^2$.
Suppose, that $\theta \ll 1$, so that one may neglect terms of order
$\theta^2$.
Substituting $e$ and $e'$ instead of $Q$ and $Q'$ respectively, one obtains
{}from (\ref{E}) and (\ref{Q}),
\begin{eqnarray}\label{m}
m' & = & m + \kappa^{-{1\over2}}\theta\,e\,;\\
\label{e}
e' & = & e + \kappa^{{1\over2}}\theta\,m\,.
\end{eqnarray}
Here
\begin{equation}\label{kappa}
\kappa \,=\, {{q^2\ell^2 c^2}\over{\hbar^2}}\,.
\end{equation}
One may interpret $\Delta m = \kappa^{-{1\over2}}\theta\,e$ and
$\Delta e = \kappa^{{1\over2}}\theta\,m$ as contributions of electrical
charge, $e$, and mass, $m$, to the observable mass, $m'$, and electrical
charge, $e'$, respectively.

3. Keeping only leading terms of the action's
expansion in $\ell$ \cite{0202024}, one would obtain standard equations for
Coulomb's electrostatics and Newton's gravity in nonrelativistic case.
Consider two electrical charges, $(e, m)$ and $(Q, M)$. Then, observable
quantities are $(e', m')$ and $(Q', M')$. Here $Q'$ and $M'$ are expressed
through $Q$ and $M$ similarly to (\ref{e}) and (\ref{m}).
Then, one obtains for the Newton and Coulomb potential energy of two
particles, separated at distance $r$,
\begin{equation}\label{N'C'}
V(r) \,=\, {{e'Q' - k m' M'}\over r}\,.
\end{equation}
Here $k$ is the Newton constant of gravitational interaction.
Substituting `unprimed' variables, one obtains for the potential energy,
\begin{equation}\label{NC}
V(r) \,=\, {{eQ - k m M + \sigma (eM + mQ)}\over r}\,,
\end{equation}
where
\begin{equation}\label{sigma}
\sigma \,=\, \theta \left(\sqrt{\kappa} - {k\over{\sqrt{\kappa}}}\right)\,.
\end{equation}

The new effect (proportional to $\sigma$) may be responsible for interaction
between cosmic charged particles (protons, electrons, etc.) and the Earth's
`charge', $Q_E = \sigma M_E$. Here $M_E$ is the Earth's mass.
Assuming that $\sigma < 0$,
one obtains that Earth's mass contributes negative electric charge,
which explains, why
primary cosmic rays consist mainly from positively charged particles \cite{2}.
The same assumption explains the fairweather electric field (100
Volts per metre at surface
of the Earth) \cite{1}. Due to the same effect rotation of the Earth
leads to electric
currents, which may be responsible for Earth's magnetic field.

\end{document}